\documentclass[aps,prd,twocolumn,showpacs,superscriptaddress]{revtex4-1}
\usepackage{amsbsy}
\usepackage{amsmath}
\usepackage{amstext}
\usepackage{amsfonts, amssymb}
\usepackage{hyperref}
\usepackage{graphicx}
\usepackage{subfigure}
\usepackage{xcolor}
\usepackage{overpic}
\graphicspath{{pic/}}
\begin{document}
%\begin{CJK*}{GB}{gbsn}
%\begin{CJK*}{GBK}{song}

%\fancyhead[c]{\small Submitted to Chinese Physics C}%%~~~Vol. xx, No. x (201x) xxxxxx}
%\fancyfoot[C]{\small 010201-\thepage}

%\footnotetext[0]{Received 31 June 2015}

\title{Localization of topological charge density near $T_c$ in quenched QCD with Wilson flow }

%\author{%
%      You-Hao Zou $^{1;1)}$\email{11006067@zju.edu.cn}%
%\quad Jian-Bo Zhang $^{1;2)}$\email{jbzhang08@zju.edu.cn}%
%\quad Guang-Yi Xiong $^{1;3)}$\email{xionggy@zju.edu.cn}%

%}

\author{You-Hao Zou}
\email{11006067@zju.edu.cn}
\affiliation{Department of Physics, Zhejiang University, Zhejiang 310027, P.R. China}
\author{Jian-Bo Zhang}
\email{jbzhang08@zju.edu.cn}
\affiliation{Department of Physics, Zhejiang University, Zhejiang 310027, P.R. China}
\author{Guang-Yi Xiong}
\email{xionggy@zju.edu.cn}
\affiliation{Department of Physics, Zhejiang University, Zhejiang 310027, P.R. China}

\begin{abstract}
We smear quenched lattice QCD ensembles with lattice volume $32^3\times 8$ by using Wilson flow.
Six ensembles at temperature near the critical temperature $T_c$ corresponding
to the critical inverse coupling $\beta_c=6.06173(49)$ are used to investigate the localization
 of topological charge density. If the effective smearing radius of Wilson flow is large
enough, the density, size and peak of Harrington-Shepard (HS) caloron-like topological lumps of ensembles
are stable when $\beta\leq 6.050$, but start to change significantly when  $\beta\geq6.055$.
The inverse participation ratio (IPR) of topological charge density shows similar results, it
begins to increase when $\beta\geq 6.055$ and is stable when $\beta\leq 6.050$. The pseudoscalar
glueball mass is extracted from the topological charge density correlator (TCDC) of ensembles at
$T=1.19T_c,~\textrm{and }1.36T_c$, the masses are $1.915(98)\textrm{ GeV}$ and
$1.829(123)\textrm{ GeV}$ respectively, they are consistent with results
from conventional methods.

\end{abstract}

%\pacs{11.15.Ha, 12.38.-t, 12.38.Gc}
\maketitle

%\address{%
%$^1$ Department of Physics, Zhejiang University, Zhejiang 310027,
%P.R. China\\
%$^2$ Institute of High Energy Physics, Chinese Academy of Sciences, Beijing 100049, China\\
%$^3$ Theoretical Center for Science Facilities, Chinese Academy of Sciences, Beijing 100049,P.R. China \\
%$^4$ School of Physics, Peking University, Beijing 100871, P.R. China\\
%$^5$ Collaborative Innovation Center of Quantum Matter, Beijing 100871, P.R. China\\
%$^6$ School of Physics, Nankai University, Tianjin 300071, P.R. China \\
%$^7$ Institute of Theoretical Physics, Chinese Academy of Sciences,Beijing 100080, P.R. China
%}

%\begin{keywords}
%topological structure,  localization of topological charge density, pseudoscalar glueball mass, flow, HS calorons
%\end{keywords}

%\keywords{topological structure,  localization of topological charge density, pseudoscalar glueball mass, flow, HS calorons}
%\begin{pacs}
%11.15.Ha, 12.38.-t, 12.38.Gc
%\end{pacs}

%\footnotetext[0]{\hspace*{-3mm}\raisebox{0.3ex}{$\scriptstyle\copyright$}2013
%Chinese Physical Society and the Institute of High Energy Physics
%of the Chinese Academy of Sciences and the Institute
%of Modern Physics of the Chinese Academy of Sciences and IOP Publishing Ltd}%

%\linenumbers
%\begin{multicols}{2}

\section{Introduction}

Topological properties of the QCD vacuum are believed to play an important role in QCD.
For example, the topological  susceptibility  has the famous  Witten-Veneziano relation,
which can explain the U(1) anomaly and the large mass of the $\eta^\prime$ meson~\cite
{WittenVeneziano1, WittenVeneziano2, WittenVeneziano3}. The topological
structure of the QCD vacuum is related to chiral symmetry breaking and may be also
related to confinement~\cite{Witten-instanton, Instantons1996}.

A usual way to study the topological structure is investigating the localization
of topological charge density, such as BPST instantons-like localized topological lumps
 at zero temperature. Instanton is a semi-classical solution of the QCD Lagrangian in
 Euclidean space~\cite{instantons1}. Isolated instantons are zero modes of the Dirac operator.
 When these modes mix with each other they will shift away from zero modes~\cite{Instantons1996}.
 The way how they mix is important, since it is the topological structure of the QCD vacuum.
 When we use the gluonic definition for the topological charge density $q(x)$ to investigate the
 topological localized structures, such as instantons, a UV filter is needed to remove the
 short-ranged topological fluctuations and preserve the long-ranged
topological structures~\cite{lumps1, lumps2, lumps3, lumps4, lumps5}.

Since the topological structure is connected with chiral symmetry breaking and confinement,
we are interested in the behavior of topological structures when the temperature is near the
critical temperature $T_c$. The temperature in lattice QCD is given by:
\begin{equation}\label{eq:temperature001}
  T = \frac{1}{N_t a_t},
\end{equation}
in which $a_t$ is the lattice spacing in the temporal direction, and $N_t$ is the temporal
lattice size. Therefore we can change $N_t$ or $a_t$ to vary the temperature $T$.
If we change $N_t$, because $N_t$ cannot be too large the temperature will be changed
coarsely. Thus we cannot get different ensembles with small variation of temperature near $T_c$.
Therefore we will vary the temperature by changing $a_t$, which means that we will generate
different temperature ensembles  by slightly varying the inverse coupling $\beta$. The conventional
UV filters like cooling, smoothing and smearing~\cite{BERG_cooling, APE, HYP, stout-link, oi-stout-link} lead
to different smearing effects when the ensembles have different lattice spacings, even though the
parameters are set to be the same. So we will use the gradient flow, which provides a general
energy scale. Its effective smearing radius $\lambda=\sqrt{8t}$~\cite{wilsonflow1}, where $t$
is the flow time.  Recent works~\cite{cooling_and_flow1, cooling_and_flow2, cooling_and_flow3} show
that the gradient flow is consistent with standard cooling, therefore like using cooling we can also use the
gradient flow to study topological structures. Then we can compare the topological structure of different ensembles
and avoid the different smearing effects.

In our work, we used the Harrington-Shepard (HS) caloron solutions~\cite{HScaloron1} to filter the localized topological
lumps, which is the generalized form of BPST instantons at finite temperature with periodic boundary
condition at the temporal direction. We also used the inverse participation ratio (IPR)~\cite{IPR1}
to investigate the topological localization. The IPR is defined by:
\begin{equation}\label{eq:IPR1}
  \textrm{IPR}=V\frac{\sum_x |q(x)|^2}{(\sum_x |q(x)|)^2},
\end{equation}
in which $q(x)$ is the topological charge density. In this work we use the gluonic definition for $q(x)$:
\begin{equation}\label{eq:qx001}
  q(x)=\frac{1}{32\pi^2}\epsilon_{\mu\nu\rho\sigma}\textrm{tr}_{C}[F_{\mu\nu}(x)F_{\rho\sigma}(x)],
\end{equation}
in which $\epsilon_{\mu\nu\rho\sigma}$ is the Levi-Civita symbol, $\textrm{tr}_{C}$ is the trace running
over the color space, and the field tensor $F_{\mu\nu}$ is defined by:
\begin{equation}\label{eq:fmn}
\begin{split}
   F_{\mu\nu}(x)= &-\frac{i}{2}(C_{\mu\nu}(x)-C^\dagger_{\mu\nu}(x))-\frac{1}{3}\textrm{ReTr}_{C}(-\frac{i}{2}\\
   &(C_{\mu\nu}(x)-C^\dagger_{\mu\nu}(x))),
\end{split}
\end{equation}
in which $C_{\mu\nu}(x)$ is the average of the four plaquettes on the $\mu-\nu$ plane. When
all topological charges focus on one lattice site $\textrm{IPR}=V$, IPR would decrease if the topological
charge density becomes more delocalized. Finally it will equal to $1$ when the topological charge density
distributes uniformly.

The topological charge density correlator(TCDC) of quenched QCD can be used to extract pseudoscalar
glueball masses at zero temperature with Wilson flow~\cite{gbmass_fit1}. In our work, we extracted
the pseudoscalar glueball mass from TCDC at finite temperature with Wilson flow. The results are
compared with those from Ref.~\cite{gbmass_fit2}. Unlike conventional methods, this method
doesn't need large lattice size in the temporal direction to do fitting, which is hard to
be satisfied in ensembles at finite temperature especially at high temperatures.

\section{Locating the HS caloron-like topological lumps}

\subsection{Find the critical inverse coupling $\beta_c$}
First, we need to find the critical temperature $T_c$. In other words we need to determine the critical inverse
coupling $\beta_c$. We use pure gauge ensembles that have lattice size $32^3\times 8$ in our
work. We use the susceptibility $\chi_P$ of the Polyakov loop to find
$\beta_c$. $\chi_P$ is defined as
\begin{equation}\label{eq:Polyakov_susc1}
  \chi_P=\langle \Theta^2\rangle - \langle \Theta\rangle^2,
\end{equation}
in which $\Theta$ is the $Z(3)$ rotated Polyakov loop:
\begin{equation}\label{eq:Polyakov_susc2}
  \Theta=
  \begin{cases}
  \textrm{Re} P \exp[-2i\pi/3]&,\quad \arg P\in[\pi/3,\pi), \\
  \textrm{Re} P  &, \quad \arg P\in[-\pi/3,\pi/3), \\
  \textrm{Re} P \exp[2i\pi /3] &,\quad \arg P\in[-\pi,-\pi/3),
  \end{cases}
\end{equation}
where $P$ is the usual Polyakov loop of each configuration.

In Table~\ref{tab:beta_c} the 6 ensembles we used to find $\beta_c$ are listed. The lattice
size is $32^3\times 8$. We expect that the finite volume effects are negligible. The lattice spacing $a$ is
found by using~\cite{Wilson_Scale}
\begin{equation}\label{eq:scale}
\begin{split}
  a=&r_0\exp(-1.6804-1.7331(\beta-6)+ \\
     & 0.7849(\beta-6)^2-0.4428(\beta-6)^3),
\end{split}
\end{equation}
where $r_0$ is set to be $0.5\textrm{fm}$ from Ref.~\cite{Sommer_R0}. Obviously Table~\ref{tab:beta_c} shows that
$\beta_c$ is near $6.060$. The critical inverse coupling $\beta_c$ is obtained by interpolating to the location
where $\chi_P$ is maximum. We use a B-spline interpolation and obtain $\beta_c=6.06173(49)$,
which is compatible with $\beta_c=6.06239(38)$ in Ref.~\cite{criticalTc}.

\begin{table}[!htb]
\caption{\label{tab:beta_c}The quenched ensembles of Wilson action in this work. The lattice size is $32^3\times 8$.
10000 sweeps were done before thermalization. Each configuration is separated by 10 sweeps. Each sweep includes
5 times quasi heat-bath and 5 steps of leapfrog.\\}
\normalsize
\begin{tabular}{|c|c|c|c|c|}
  \hline
  $\beta$ & $N_{cnfg}$ &$\chi_P$ & $a$ \\
  \hline
  6.045 & 2000 & $3.02(38)\times 10^{-4}$ &0.0863fm \\
  \hline
  6.050 & 2000 & $4.92(14)\times 10^{-4}$ &0.0856fm\\
  \hline
  6.055 & 2000 & $7.67(23)\times 10^{-4}$ &0.0849fm\\
  \hline
  6.060 & 2000 & $9.36(47)\times 10^{-4}$ &0.0842fm\\
  \hline
  6.065 & 2000 & $8.15(19)\times 10^{-4}$ & 0.0835fm\\
  \hline
  6.070 & 2000 & $ 5.82(18)\times 10^{-4}$ &0.0828fm\\
  \hline
\end{tabular}
\end{table}

\subsection{HS caloron-like topological lumps }

In this paper we use the HS caloron solutions to filter the localized topological charge density lumps.
The localized topological lumps are defined by sites that have maximum absolute value of
$q(x_c)$ in a $3^4$ hypercube centered at site $x_c$. The center $x_c$ is also mentioned as peak.
After applying the HS caloron filters in the following, we can get calorons-like topological lumps.

In SU(2) gauge theory at temperature $T$, HS caloron solution of gauge field $A_\mu(x)$ has
the exact form as~\cite{HScaloron1}
 \begin{widetext}
\begin{equation}\label{eq:caloron001}
\begin{split}
  A_\mu(x)& =A^a_\mu(x)T^a,\ \textrm{$T^a$ is the generators for SU(2)}, \\
  A^a_\mu(x)=\eta^{(\pm)}_{a\mu\nu}\partial_\nu \ln \Phi(x),\ \Phi(x)&
  =1+\frac{\pi \rho^2}{|\vec{x}-\vec{x}_c|/T} \frac{\sinh(2T\pi|\vec{x}-\vec{x}_c|)}
  {\cosh(2T\pi|\vec{x}-\vec{x}_c|)-\cos(2T\pi(x_4-x_{c4}))},
\end{split}
\end{equation}
\end{widetext}
where $x_c$ is the center of a HS caloron, $\rho$ is the size of a HS caloron. It satisfies the
(anti-)self-dual condition $F_{\mu\nu}=\pm \tilde F_{\mu\nu},\ \tilde F_{\mu\nu}=\frac{1}{2}
\epsilon_{\mu\nu\rho\sigma}F_{\rho\sigma}$,
$\eta_{a\mu\nu}^{(\pm)}$ is the 't Hooft symbol:
 \begin{equation}\label{eq:tHooftSymbol}
\begin{split}
 & \eta_{a\mu\nu}^{(\pm)}=\epsilon_{a\mu\nu},~\mu,\nu=1,2,3,\\
 & \eta_{a4\nu}^{(\pm)}=-\eta_{a\nu4}^{(\pm)}=\pm \delta_{a\nu},~\eta_{a44}^{(\pm)}=0.
\end{split}
\end{equation}
When the temperature $T\rightarrow 0$, it approaches the BPST instanton solution $\Phi(x)
\rightarrow 1+ \frac{\rho^2}{(x-x_c)^2}$~\cite{instantons1}. Similar things happen when we constrain
our study at the region $|x-x_c| \ll 1/T=N_ta_t$. Therefore when we use the center and
its 8 closest neighbour sites on the lattice to filter the topological lumps with HS calorons,
we can just use the BPST instanton solution to approximate the HS caloron solution in SU(3):
\begin{equation}\label{eq:BPST001}
\begin{split}
  &A^{a(BPST)}_\mu(x)=2R^{a\alpha}\eta_{\alpha\mu\nu}^{(\pm)}\frac{(x-x_c)_\nu}{(x-x_c)^2}
  \frac{1}{1+\frac{(x-x_c)^2}{\rho^2}}, \\
  &a=1,2,...,8,~\alpha=1,2,3,
\end{split}
\end{equation}
where $R^{a\alpha}$ represents the color rotations embedding the SU(2) BPST instantons into SU(3).

The topological charge density near the center of an isolated instanton approximates
\begin{equation}\label{eq:qx002}
  q^{BPST}(x)=\pm \frac{6}{\pi^2\rho^4}(\frac{\rho^2}{(x-x_c)^2+\rho^2})^4,
\end{equation}
where the "$+$" sign is for instanton, "$-$" for anti-instanton. Then at the center
\begin{equation}\label{eq:qx003}
  q^{BPST}(x_c)=\pm \frac{6}{\pi^2\rho^4}.
\end{equation}

Therefore we can get the relation
\begin{equation}\label{eq:qx004}
  \frac{q(x)}{q(x_c)}=(\frac{\rho^2}{(x-x_c)^2+\rho^2})^4.
\end{equation}
In this paper we use the peak and the 8 closest neighbour sites on the lattice to fit Eq.~\eqref{eq:qx004}
to get the size $\rho$.

Like in Ref.~\cite{Find_instanton}, we also use 3 filter conditions to find HS caloron-like
topological lumps:
\begin{itemize}
  \item
\begin{equation}\label{eq:filter001}
  \frac{\sqrt[4]{\frac{6}{\pi^2q(x_c)}}}{\rho} \in (1-\epsilon_R,1+\epsilon_R),
\end{equation}
which comes from Eq.~\eqref{eq:qx003}.
  \item
  \begin{equation}\label{eq:filter002}
  \frac{\sum_{|x-x_c|\le a} q(x)}{\sum_{|x-x_c|\le a} s(x)} \in (1-\epsilon_S,1+\epsilon_S),
\end{equation}
where the normalized action density $s(x)=\frac{a^4}{8\pi^2}\sum_{\mu<\nu}
\textrm{tr}_{C}F_{\mu\nu}^2(x)$, the normalization factor $8\pi^2$ comes from the
action of a single HS caloron $S=\frac{g^2}{\pi^2}|Q|$ with
$Q=\int d^4x q(x)$.
  \item To avoid double countings of two peaks of a single but distorted HS caloron,
  we filter peak $x_{c^\prime}$ by
  \begin{equation}\label{eq:filter003}
  \textrm{if}~ |x_c-x_{c^\prime}|<\epsilon\rho(x_c).
\end{equation}
The topological lump centering at $x_{c^\prime}$ will be filtered.
\end{itemize}

\section{Localization of topological charge density}

We use the HS calorons filter conditions and IPR to investigate the localization of
topological charge density. Ensembles in Table~\ref{tab:beta_c} would be used every ten
configurations, which means that every ensemble includes 200 configurations and each
configuration is separated by 100 sweeps. We only show the figures that result from parameters
$\epsilon_R=0.5,\ \epsilon_S=0.4,\ \epsilon=0.7,\ 1.0$. But we have used parameters varied
in the regions $\epsilon_R=0.3-0.7,\ \epsilon_S=0.2-0.6,\ \epsilon=0.7-1.0$. These results
are consistent with the discussion in the following. We choose $\epsilon_R=0.5,\ \epsilon_S=0.4$
since the results are stable around them.

The gradient flow we used is of Wilson action, which means that we use Wilson flow to
smear the gauge fields. The effective smearing radius $\lambda$ runs from 0.3fm to 0.9fm.

\begin{figure}[!htb]
\subfigure[$\beta=6.045$]{
\includegraphics[width=0.22\textwidth,height=0.17\textwidth]{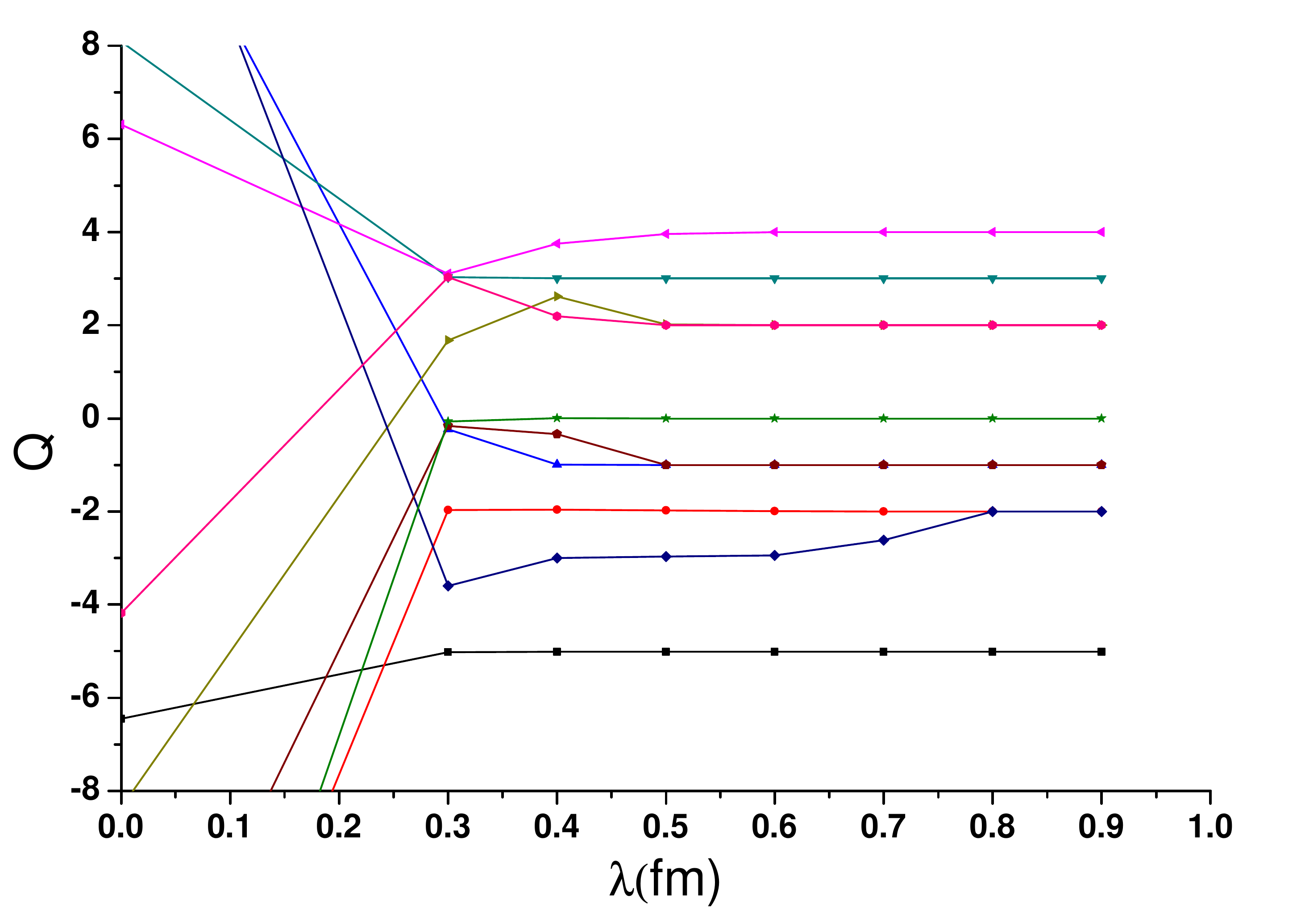}
}
\subfigure[$\beta=6.050$]{
\includegraphics[width=0.22\textwidth,height=0.17\textwidth]{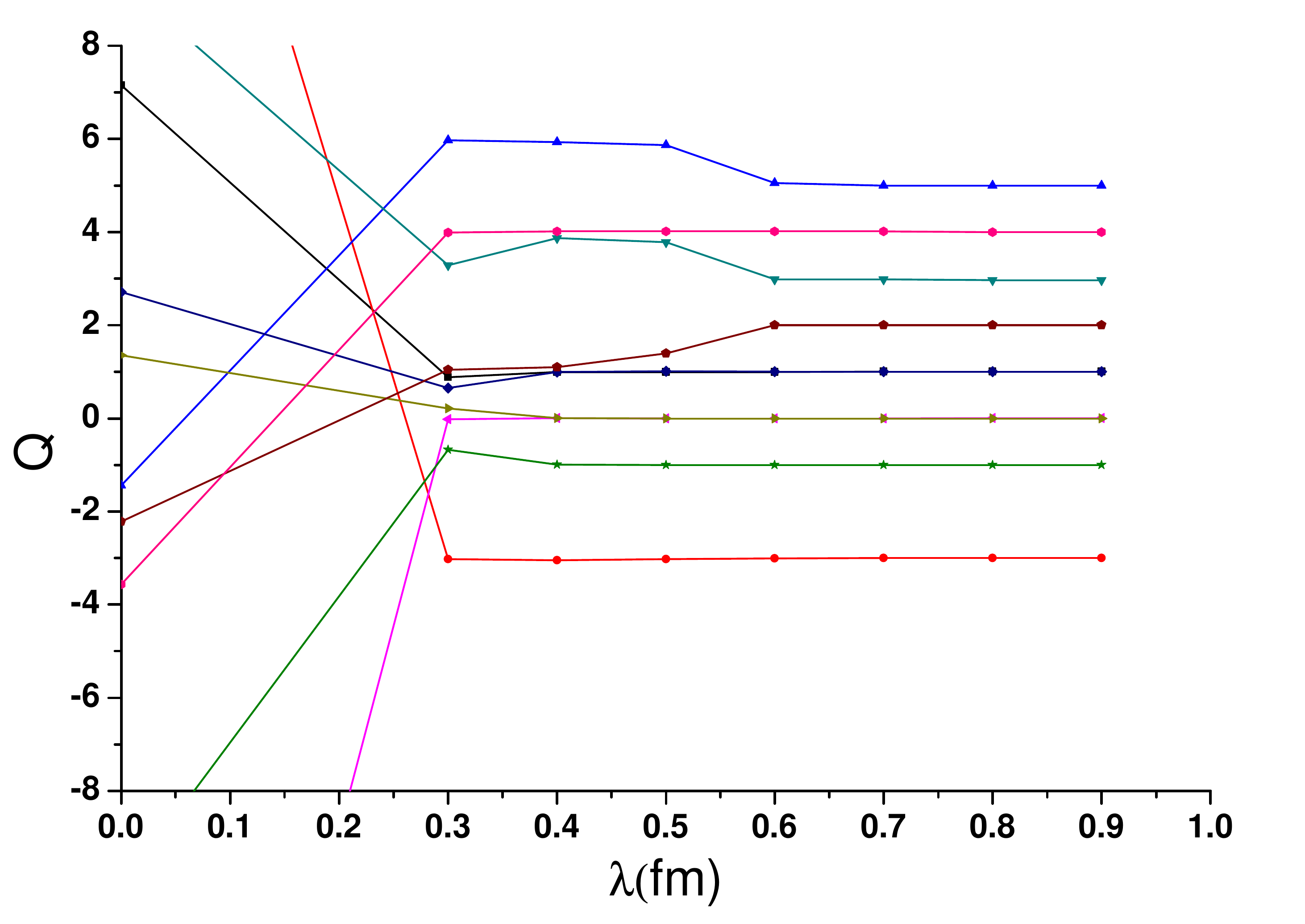}
}
\subfigure[$\beta=6.055$]{
\includegraphics[width=0.22\textwidth,height=0.17\textwidth]{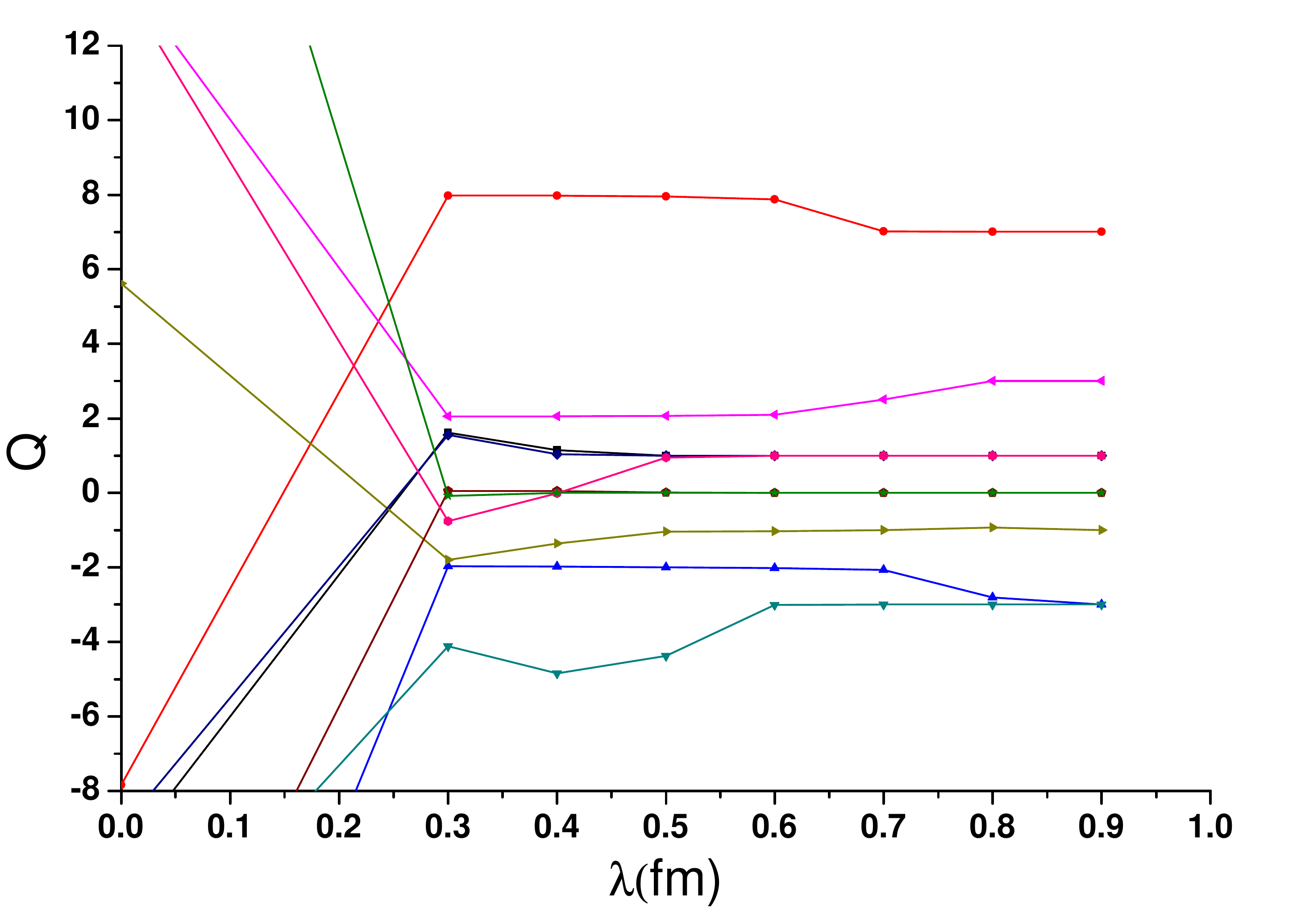}
}
\subfigure[$\beta=6.060$]{
\includegraphics[width=0.22\textwidth,height=0.17\textwidth]{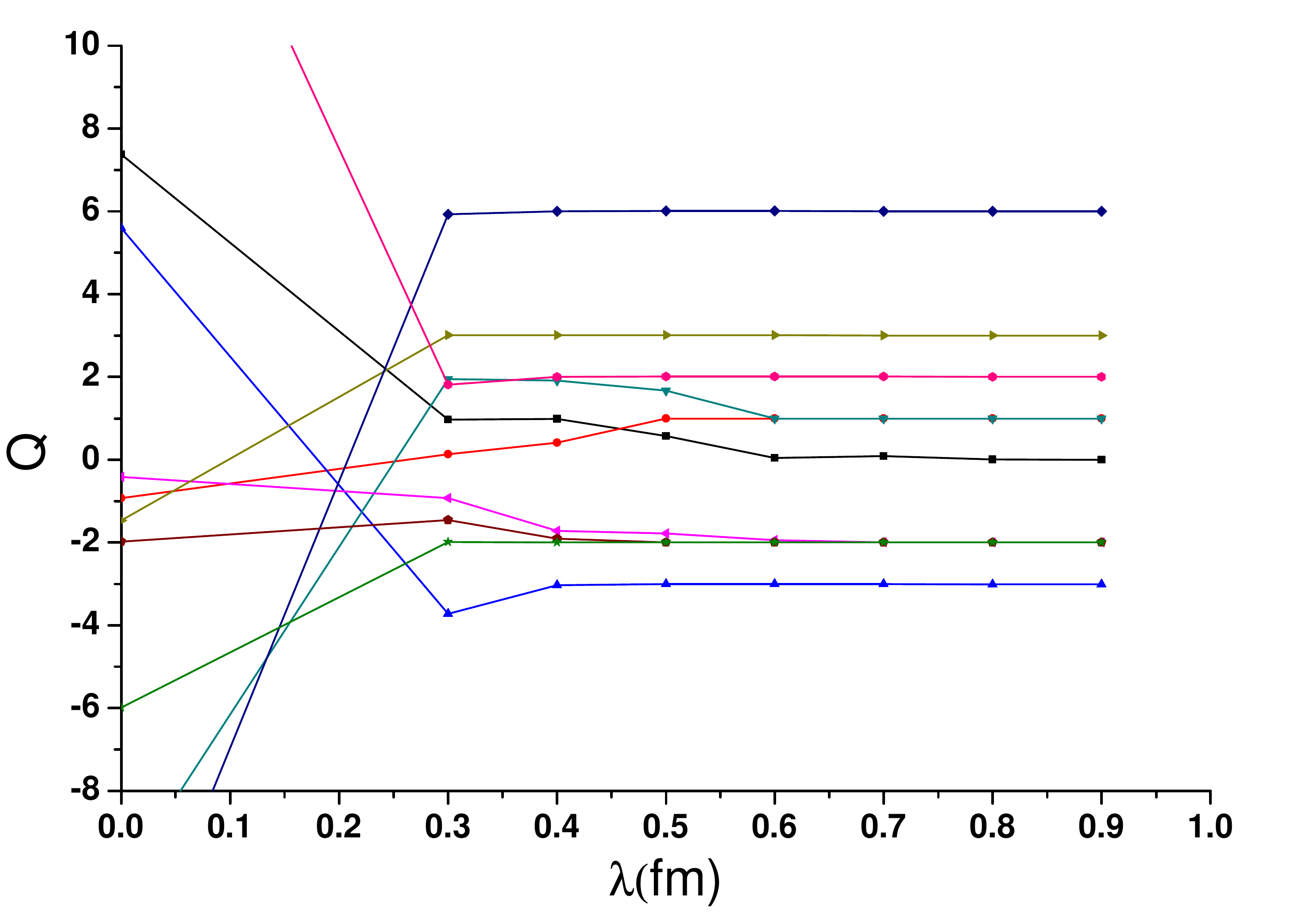}
}
\subfigure[$\beta=6.065$]{
\includegraphics[width=0.22\textwidth,height=0.17\textwidth]{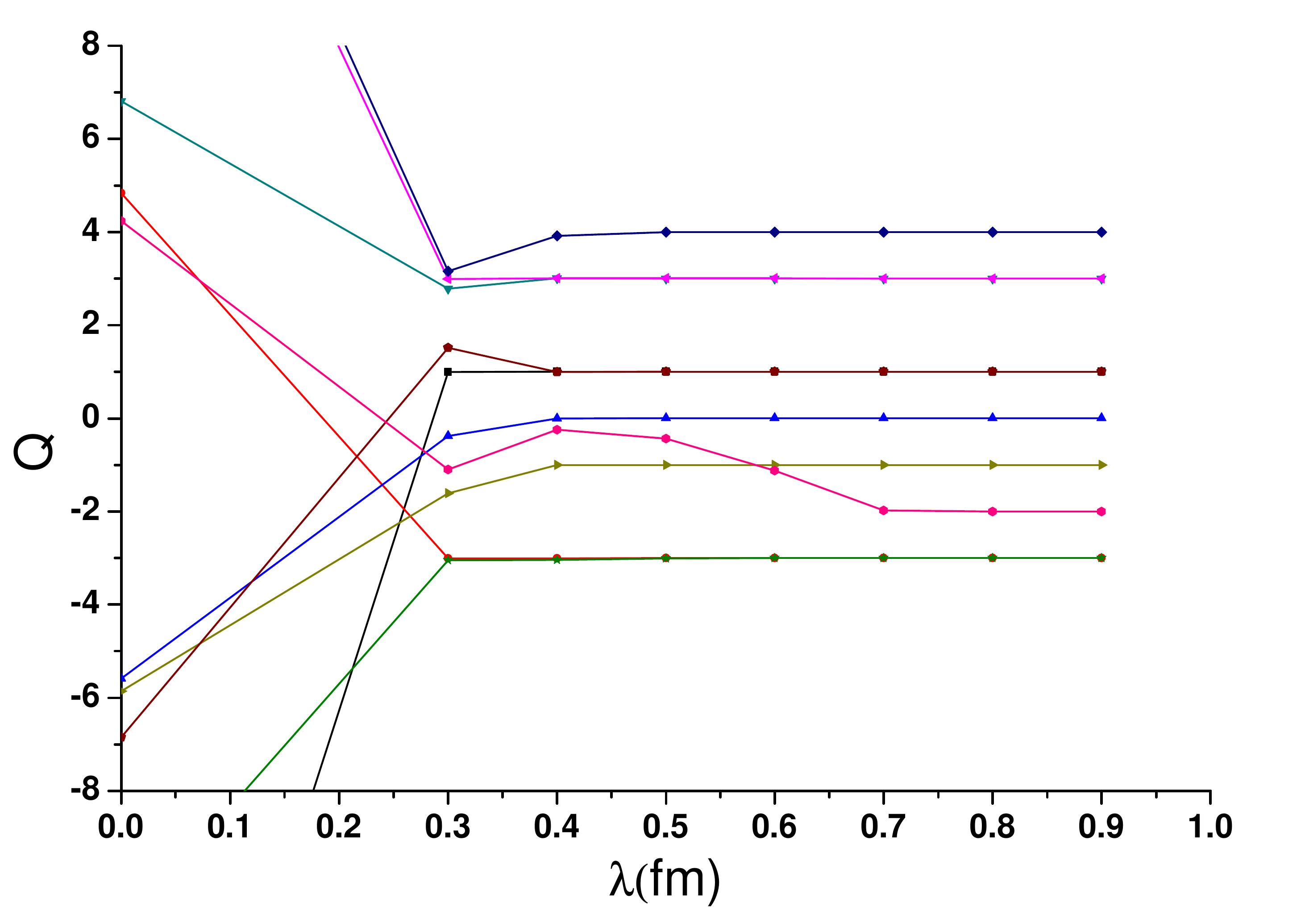}
}
\subfigure[$\beta=6.070$]{
\includegraphics[width=0.22\textwidth,height=0.17\textwidth]{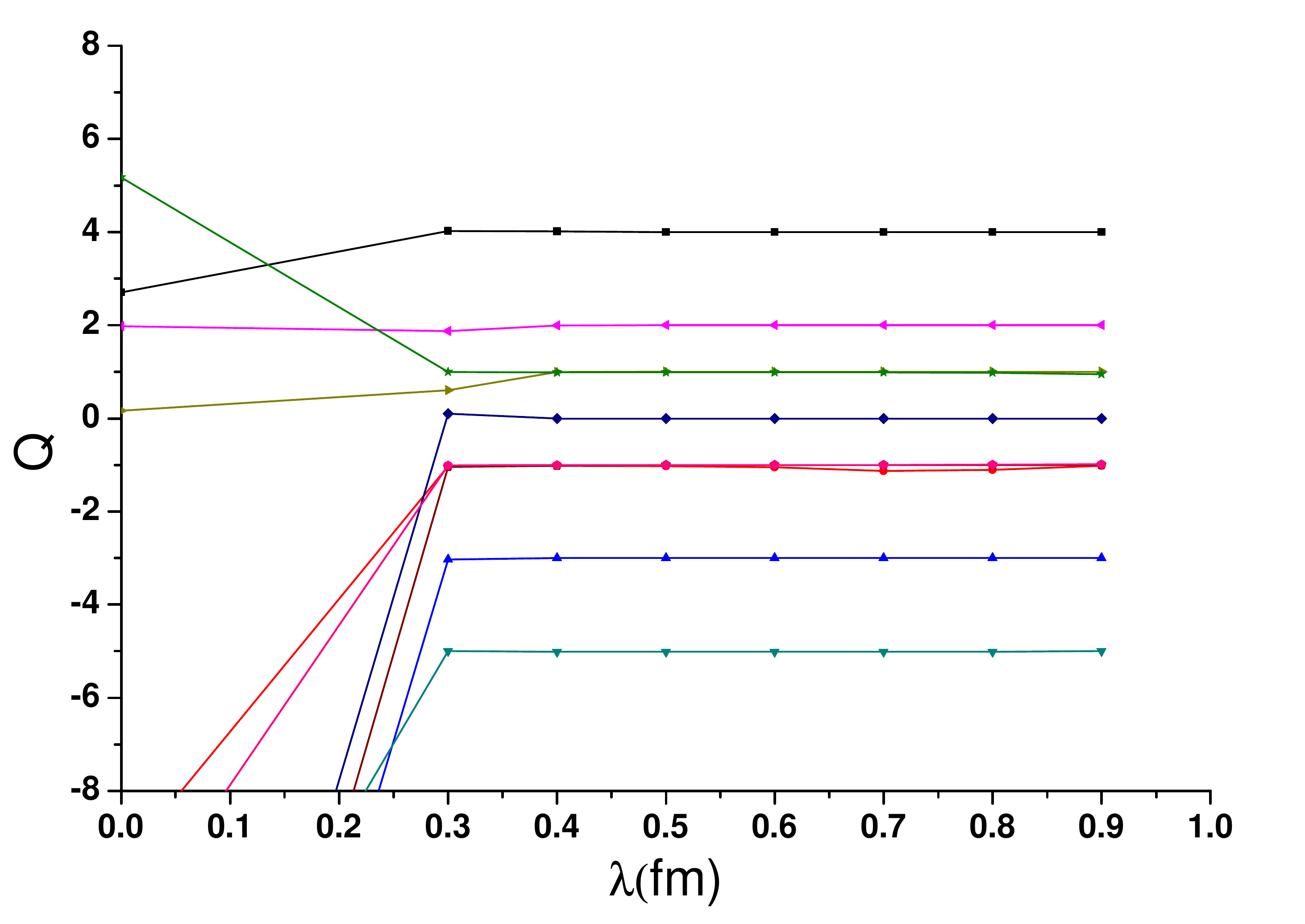}
}
\caption{\label{fig:Q_flow}The topological charges $Q$ of ten configurations versus Wilson flow,
$\lambda$ runs from 0.3fm to 0.9fm. $Q$ of original configurations have also been presented. The
topological charges approach to integers and don't drop down to the value zero during the Wilson flow. }
\end{figure}

In Fig.~\ref{fig:Q_flow}, we present the topological charges $Q$ of ten configurations
versus Wilson flow in every ensemble, the topological charges $Q$ of the original configurations have
also been presented. Obviously when $\lambda$ runs from 0.3fm to 0.9fm, the topological charges
$Q$ approach to integers. At the same time the topological charges $Q$ don't drop down to the value zero.
Therefore the long-ranged topological structures should be preserved during the Wilson flow.

\subsection{Investigating the HS caloron-like topological lumps}\label{quantities_caloron}

\begin{figure*}[!htb]
\includegraphics*[viewport=3cm 1cm 26cm 19cm, width=0.48\textwidth,height=0.38\textwidth]{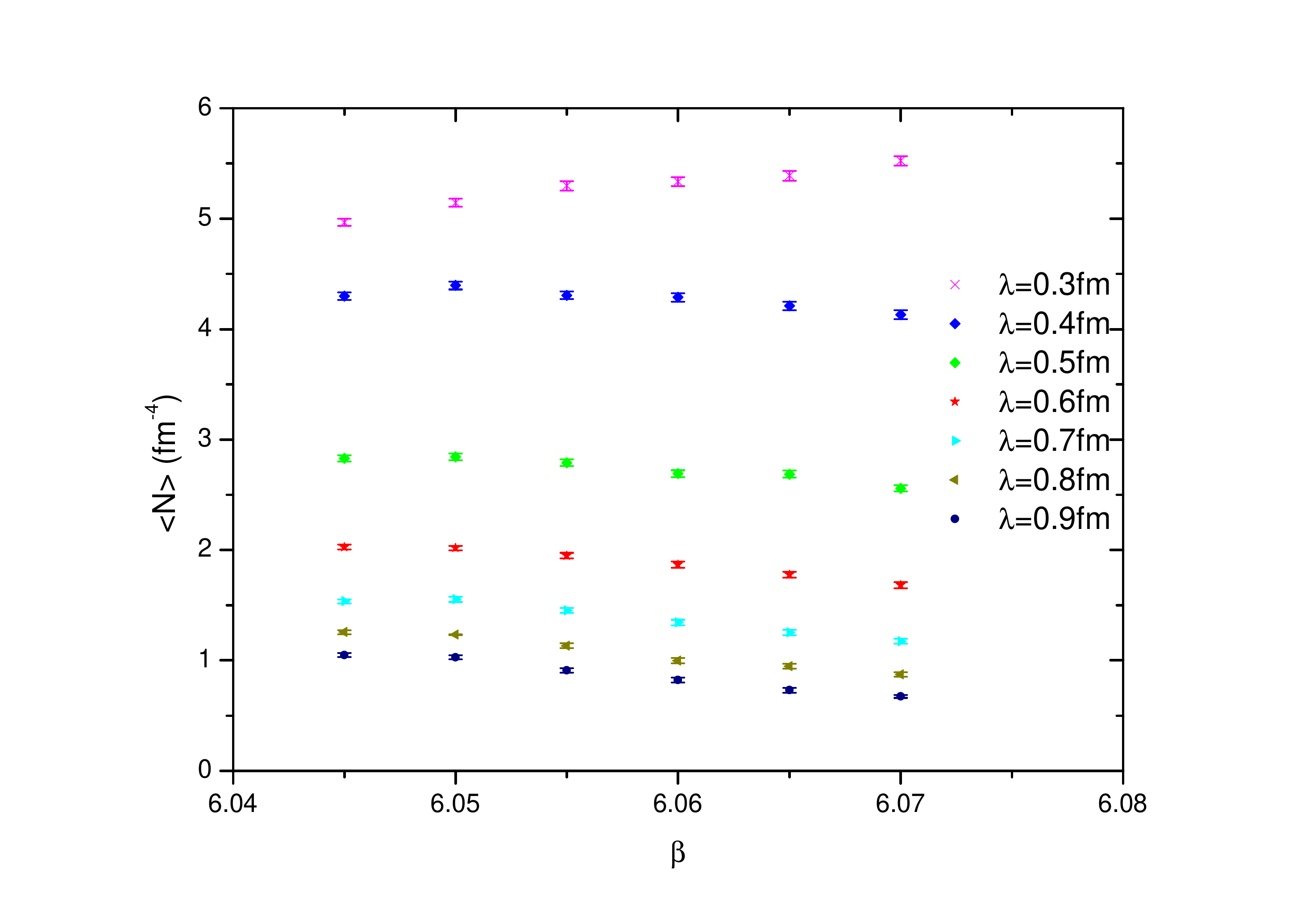}
\includegraphics*[viewport=3cm 1cm 26cm 19cm, width=0.48\textwidth,height=0.38\textwidth]{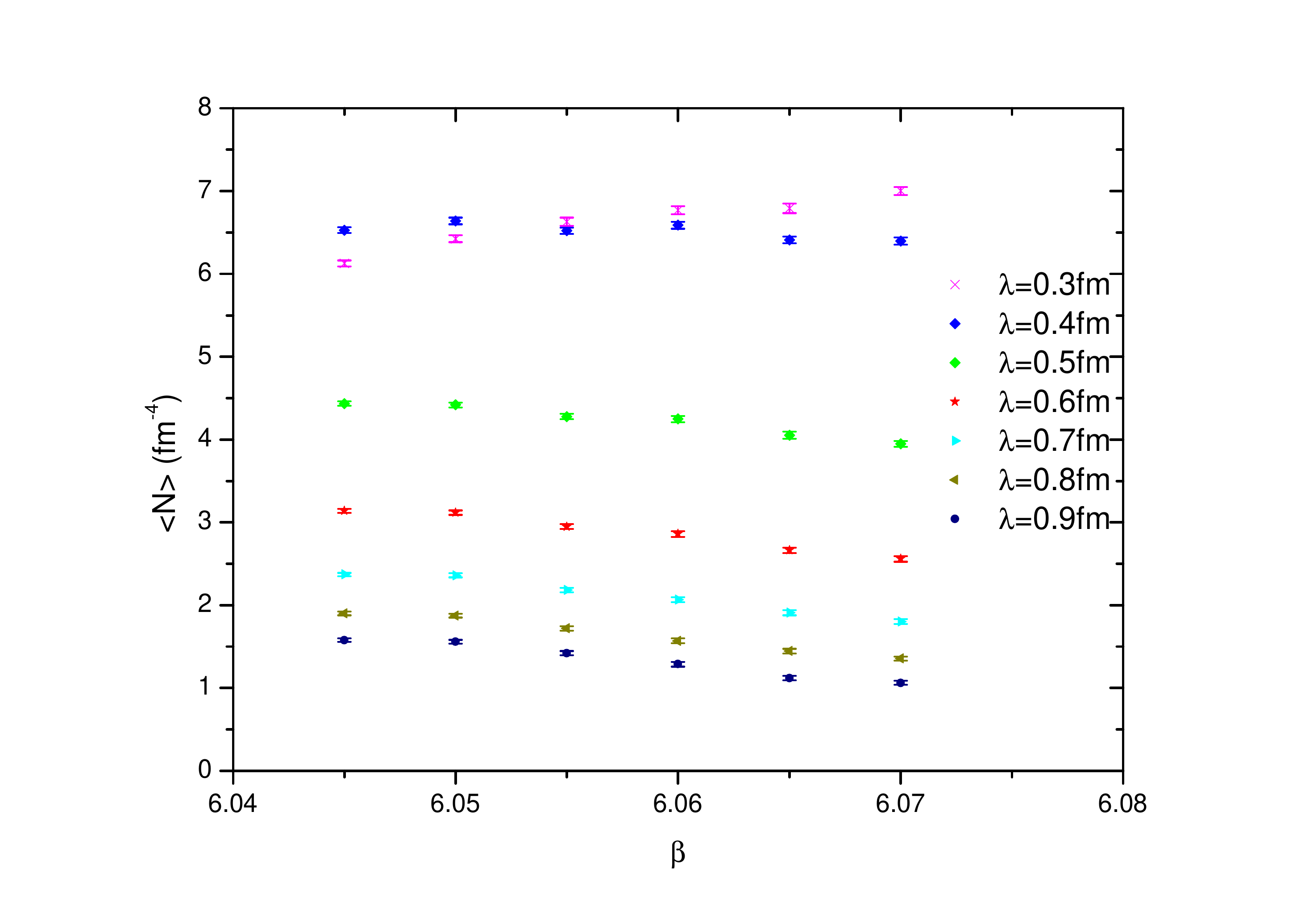}
\includegraphics*[viewport=3cm 1cm 26cm 19cm, width=0.48\textwidth,height=0.38\textwidth]{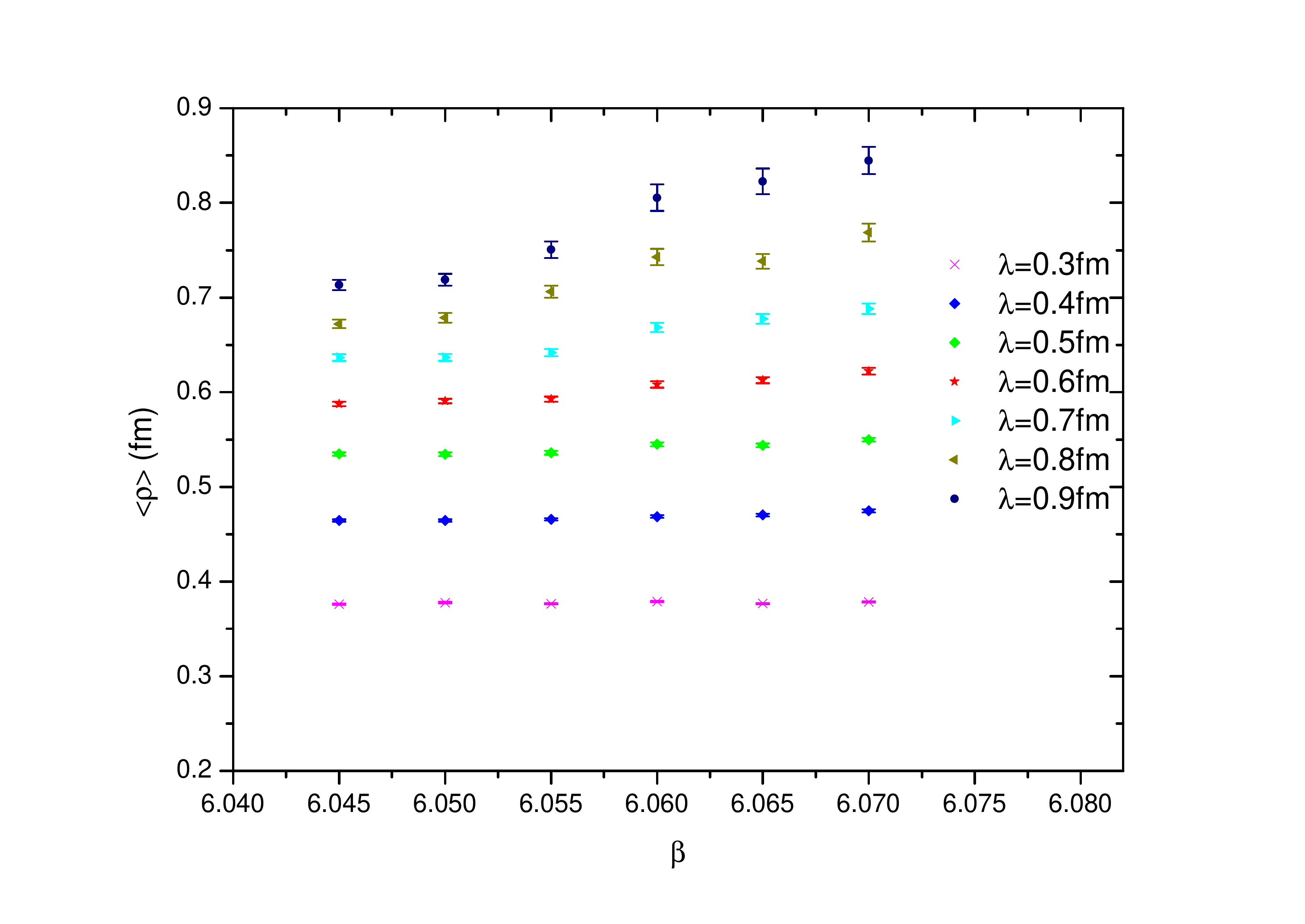}
\includegraphics*[viewport=3cm 1cm 26cm 19cm, width=0.48\textwidth,height=0.38\textwidth]{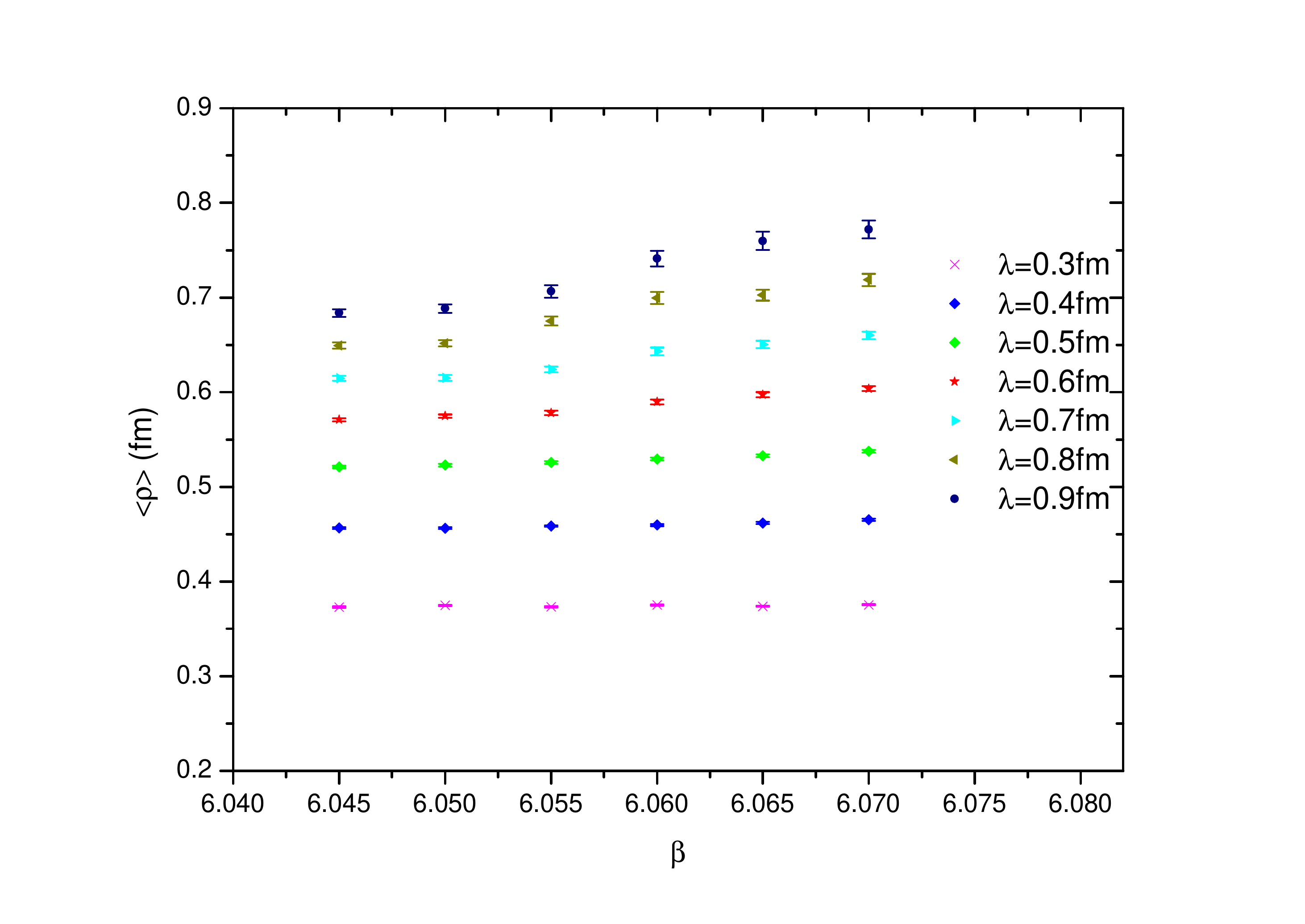}
\includegraphics*[viewport=3cm 1cm 26cm 19cm, width=0.48\textwidth,height=0.38\textwidth]{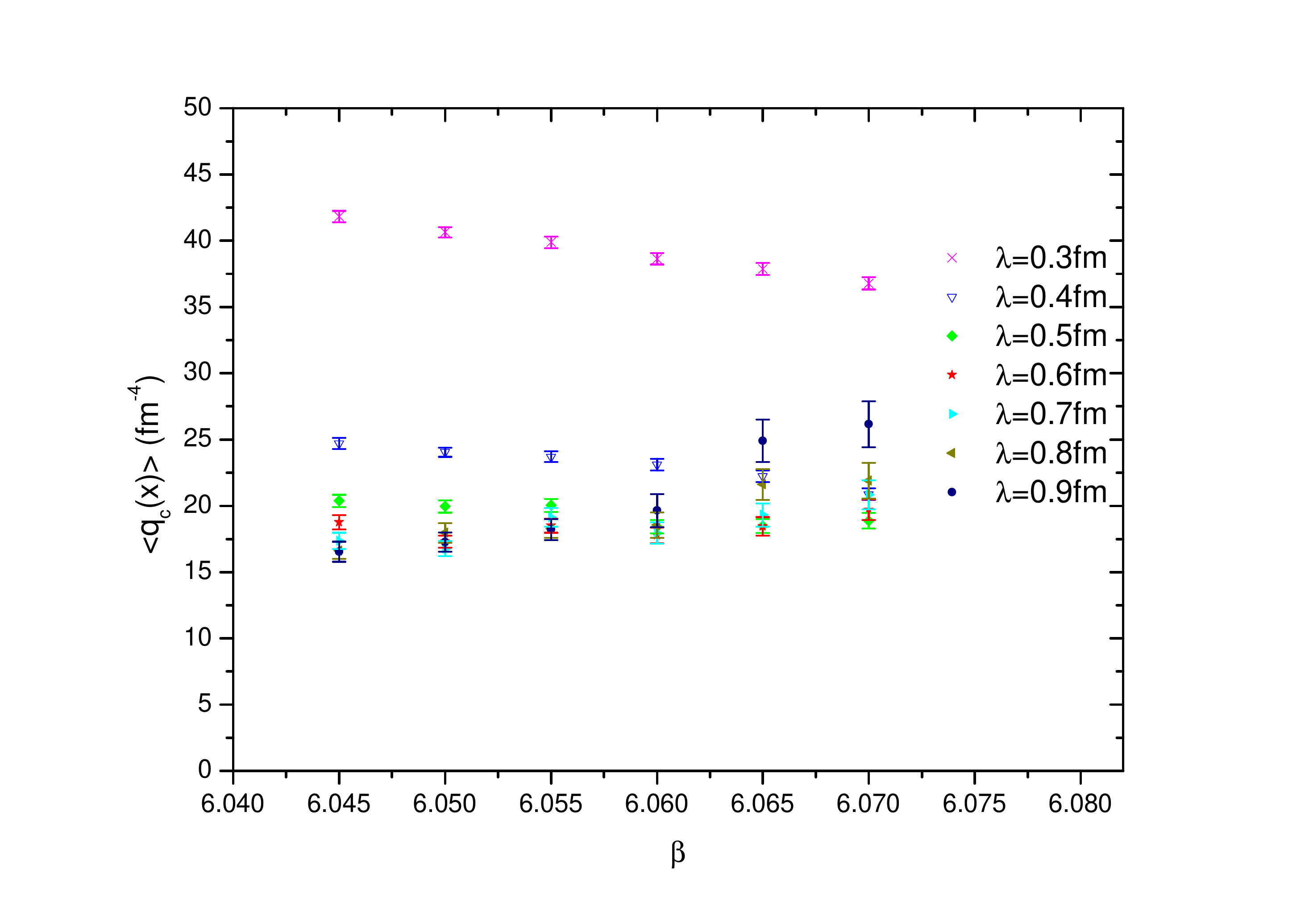}
\includegraphics*[viewport=3cm 1cm 26cm 19cm, width=0.48\textwidth,height=0.38\textwidth]{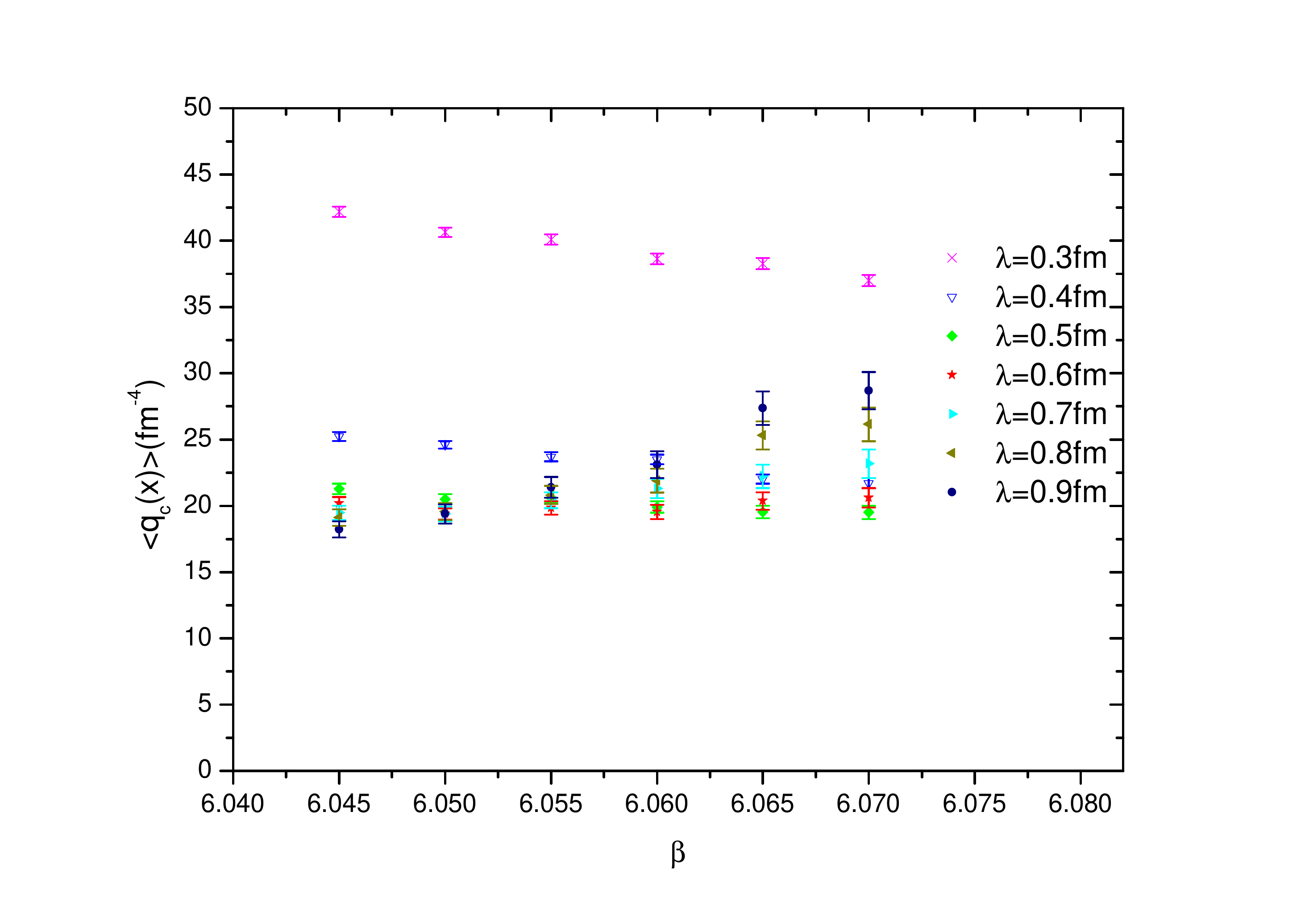}
\caption{\label{fig:calorons} The parameters setting: $\epsilon_R=0.5,\ \epsilon_S=0.4$,
left panels: $\epsilon=1.0$, right panels: $\epsilon=0.7$.
From top to bottom: the average density $\langle N\rangle $,
the average size $\langle\rho\rangle$, and the average absolute value $\langle q_c(x)\rangle$.}
\end{figure*}

In Fig.~\ref{fig:calorons}, we show the three quantities of HS caloron-like topological
lumps versus $\beta$: the average density $\langle N\rangle $, the average size
$\langle\rho\rangle$ and $\langle q_c(x)\rangle$, which is the average absolute
value of topological charge density on the peak. The three quantities with different
effective smearing radius are marked with different colors or shapes.

With the increase of the effective smearing radius $\lambda$, the average density $\langle N\rangle $
decreases monotonically, the average size $\langle\rho\rangle$ grows monotonically. Unlike
$\langle N\rangle $ and $\langle\rho\rangle$, $\langle q_c(x)\rangle$ of the ensembles at higher temperatures decreases
 at first, then becomes to increase instead as $\lambda$ increases.

The phenomena that $\langle N\rangle $ decreases monotonically and  $\langle\rho\rangle$ grows monotonically
can be expected. Since with the increase of $\lambda$, more and more small topological lumps would be
smoothed out.

When $\lambda$ is large, we find that the three quantities of HS caloron-like topological lumps are
consistent at $\beta=6.045$ and $\beta=6.050$. It indicates that the localization of topological
charge density is stable. When $\beta\geq 6.055$ , we find that the three quantities
change significantly as the the temperature increases. It means that the topological structures
have a transition point near $\beta=6.055$.

Since when $\lambda$ is small, the short-ranged fluctuations may not be suppressed enough,
we needn't pay much attention to the behaviors of the three quantities of the HS caloron-like topological
lumps at small $\lambda$.

The decrease of the average density $\langle N\rangle $ when $\beta\geq 6.055$ means that
the topological excitation is suppressed. It may explain why the topological susceptibility starts
to drop down near $T_c$~\cite{Xiong-guangyi001}.

Noting that $\frac{1}{\langle N\rangle}$, the average volume occupied by one HS caloron-like topological lump,
is always close to  $(2\langle \rho\rangle)^4$, the average volume of the HS caloron-like topological lumps.
It means that the HS caloron-like topological lumps are not sparse but dense.

Since the chiral condensate $\langle \bar{\psi}\psi \rangle \propto -\frac{\langle N\rangle^{\frac{1}{2}}}
{\langle \rho\rangle}$~\cite{Instantons1996}, the decrease of $\langle N\rangle$  and the increase of
$\langle \rho\rangle$ as the temperature increases at $\beta\geq 6.055$ indicate that the absolute value of
chiral condensate will drop down as the temperature rises. It is consistent with the fact that the chiral symmetry will restore
at high temperature.

%Someone may be concerned about that the discussions above was dependent on the HS caloron solution and
%the 3 parameters of filter we used, then it may be unreliable. Therefore in the next section we will use
%IPR tostudy the localization of topological charge density.
IPR has also been used to study the localization of $q(x)$, and conclusions from both methods are consistent.

\subsection{Average IPR versus $\beta$ with Wilson flow}

\begin{figure}[!htb]
\normalsize
\includegraphics*[viewport=3cm 1cm 27cm 20cm,width=0.45\textwidth,height=0.35\textwidth]{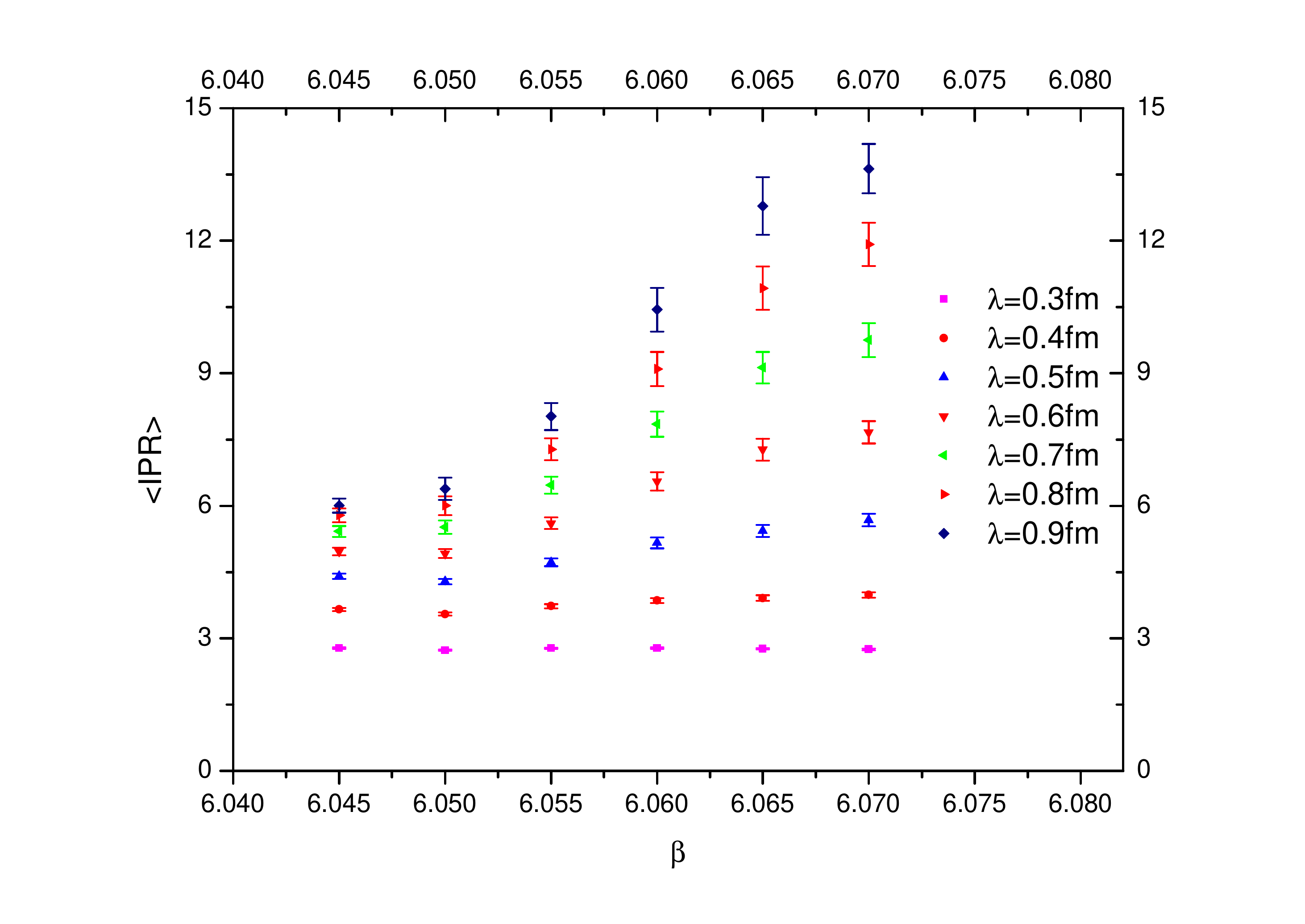}
\caption{\label{fig:IPR} $\langle \textrm{IPR}\rangle$ of topological charge density
with Wilson flow versus the inverse coupling $\beta$, the effective smearing radius
$\lambda$ of Wilson flow runs from $0.3\textrm{fm}$ to $0.9\textrm{fm}$.}
\end{figure}

In Fig.~\ref{fig:IPR} we show the average inverse participation ratio $\langle \textrm{IPR}\rangle$
versus $\beta$ with Wilson flow. Theoretically, when a certain structure is embedded in a finite $4D$ space
discretized by lattice spacing $a$,  the IPR of the structure
obeys $\textrm{IPR} \sim a^{4-d}$ as $a\rightarrow 0$~\cite{IPR1}, where $d$ denotes the dimension of the structure.
But the dependence of $\textrm{IPR}$ on the volume of the finite $4D$ space is small~\cite{IPR1}. However, when we
use gradient flow to smear the configurations in a space discretized with different lattice spacings, the average
IPR of $q(x)$ with same $\lambda$ would be almost the same if $\lambda$ is large enough, only mild scaling violation is
found~\cite{gbmass_fit1}. Therefore, any manifest differences of $\langle\textrm{IPR}\rangle$ of $q(x)$
among different temperatures can't result from the lattice discretization with different lattice spacings. The manifest
differences can only result from the different localizations of topological charge density at different temperatures.

In Fig.~\ref{fig:IPR} we find that when $\lambda$ is large, $\langle\textrm{IPR}\rangle$ increases as $\beta$ increases
when $\beta \geq 6.055$. It is just the same transition point that we found in Sect.~\ref{quantities_caloron}. Obviously,
this behaviour of $\langle \textrm{IPR}\rangle$  should come from the fact that the topological
localization was enhanced by the increase of temperature. The ensembles at $\beta=6.045$ and $\beta=6.050$ have
$\langle \textrm{IPR}\rangle$ compatible for all used $\lambda$. It means that the localization of $q(x)$ hasn't changed yet
when $\beta \leq 6.050$, just like the behaviours of the three quantities of HS caloron-like topological
lumps in Fig.~\ref{fig:calorons}.

By using the two different methods, we get the conclusion that the localization of topological charge density near
$T_c$ doesn't change when $\beta \leq 6.050$, and starts to change significantly when $\beta \geq 6.055$ . %This conclusion is consistent with that from HS calorons.

\section{Extracting the pseudoscalar glueball mass from the TCDC at high temperature}

The topological charge density correlator (TCDC) is defined by
\begin{equation}
C_{qq}(r)=\langle q(x)q(y)\rangle,~r=|x-y|.
\end{equation}
In the negative tail region of the TCDC, it can be approximated by the pseudoscalar propagator~\cite{Cqq_mass}
\begin{equation*}
  \langle q(x)q(y)\rangle = \frac{m}{4\pi^{2}r} K_{1}(mr),~~r=\mid x-y \mid ,
\end{equation*}
where $K_{1}(z)$ is the modified Bessel function, it has the asymptotic form as
\begin{equation}\label{eq:K1func}
  K_1(z) ~\underset{{\rm large}~z}{\sim} ~e^{-z} ~ \sqrt{\frac{\pi}{2z}}~ \left[1 + \frac{3}{8z}  \right ].
\end{equation}
Thus we can extract the mass of pseudoscalar particle by fitting Eq.~\eqref{eq:K1func} at zero
temperature~\cite{gbmass_fit1, eta_mass2015, myjob2017}.

We may also use Eq.~\eqref{eq:K1func} to extract the pseudoscalar glueball mass from TCDC at
finite temperature in quenched lattice QCD, the mass $m$ and amplitude are set to be two
free parameters in the fitting procedure. The procedure has been applied to the two ensembles
in Table~\ref{tab:wflow_cnfg2}. The effective smearing radius $\lambda$ of Wilson flow runs
from $0.12\textrm{fm}$ to $0.20\textrm{fm}$, each ensemble includes 500 configurations.

\begin{table}[!htb]
\caption{\label{tab:wflow_cnfg2} The quenched ensembles of Wilson action in this work. The lattice size is $32^3\times 8$.
10000 sweeps were done before thermalization. Each configuration is separated by 100 sweeps. Each sweep includes
5 times quasi heat-bath and 5 steps of leapfrog.\\ }
\begin{tabular}{|c|c|c|}
  \hline
  $\beta$ & 6.170 & 6.236  \\
  \hline
  $N_{cnfg}$ & 500 & 500  \\
  \hline
   $T$ &   $1.19T_c$ & $1.36T_c$\\
  \hline
\end{tabular}
\end{table}

\begin{figure*}[!htb]
\centering
\includegraphics*[viewport=3cm 1cm 26.5cm 19cm, width=0.48\textwidth,height=0.4\textwidth]{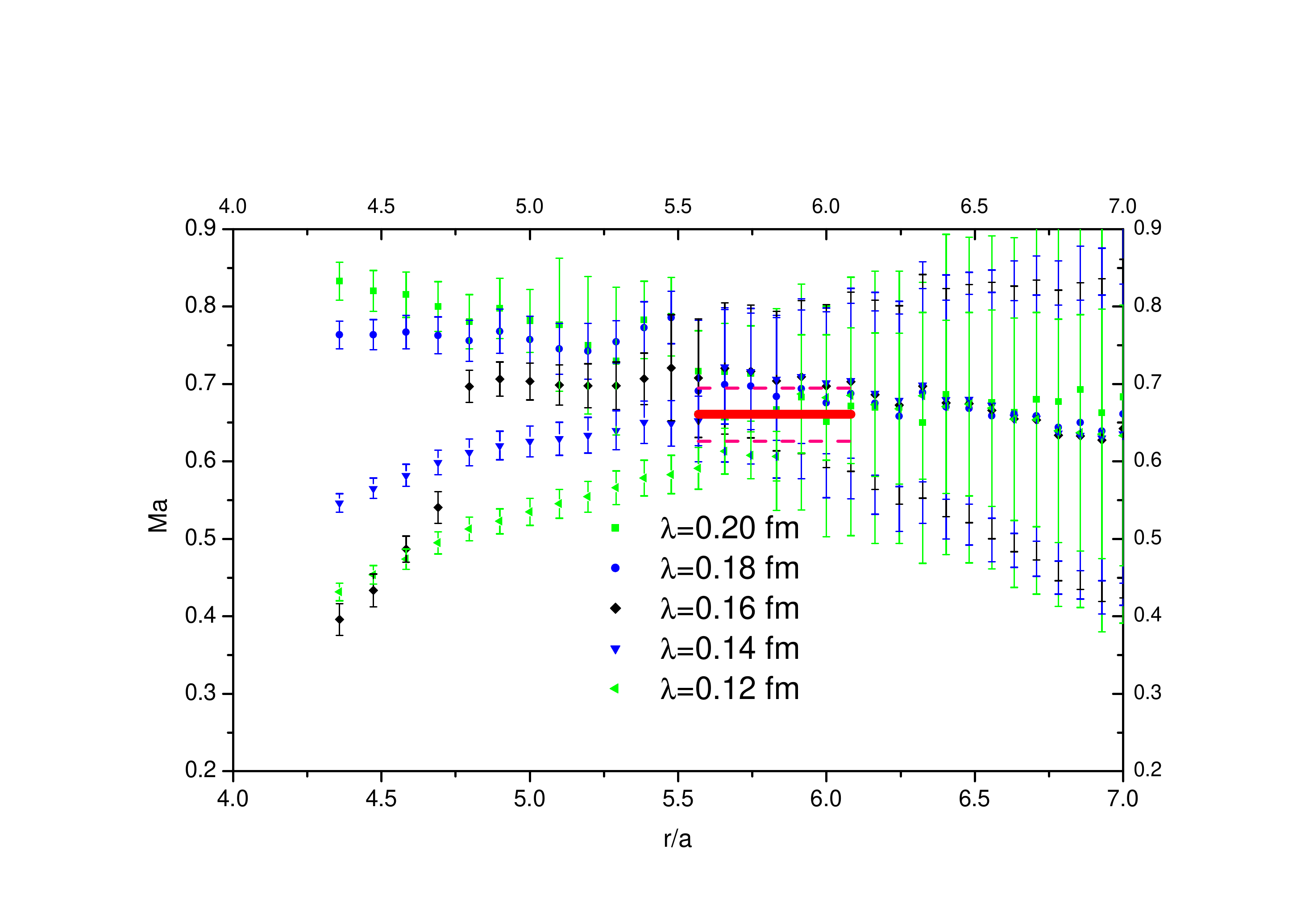}
\includegraphics*[viewport=3cm 1cm 26.5cm 19cm, width=0.48\textwidth,height=0.4\textwidth]{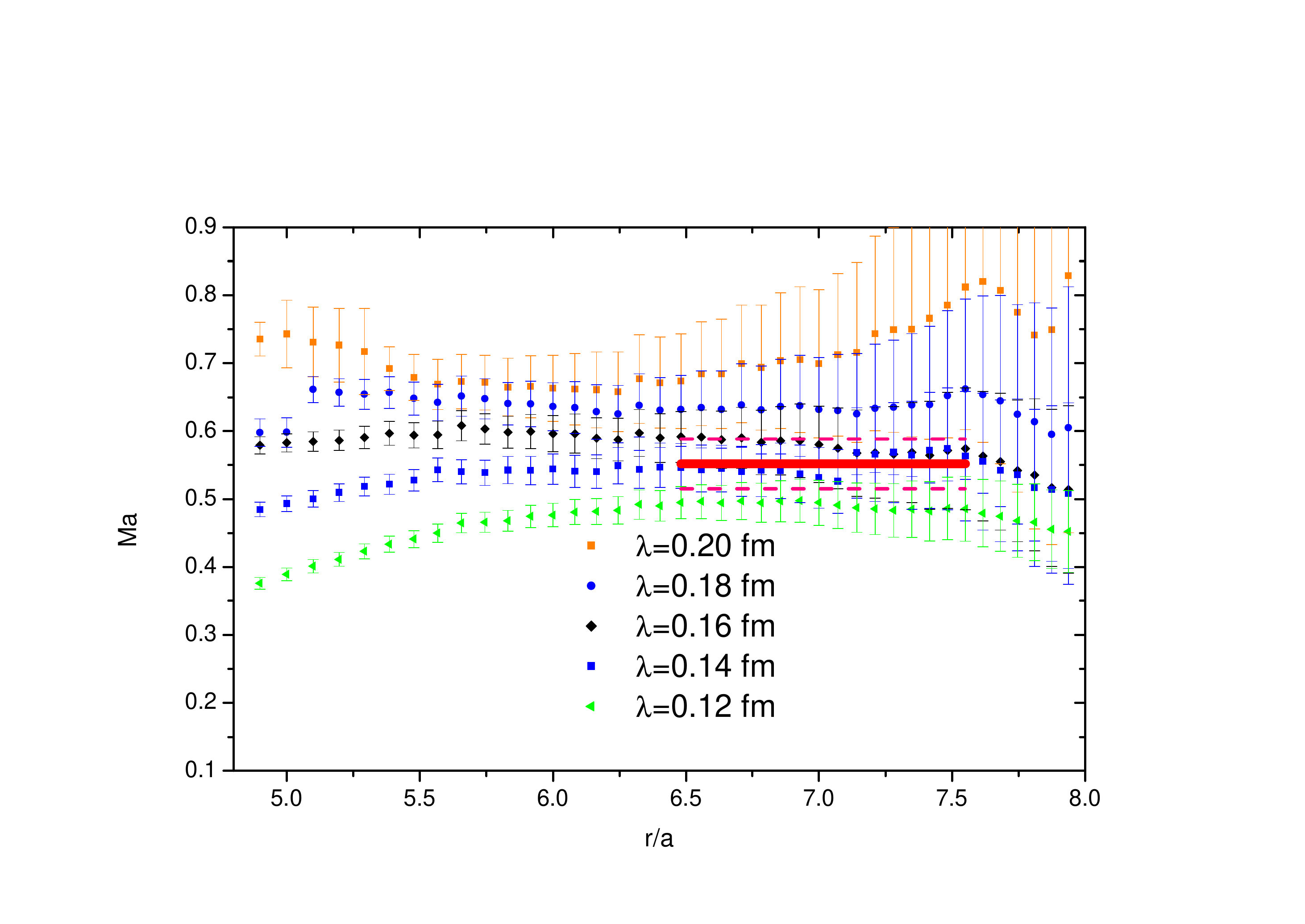}
\caption{\label{fig:mass001} The pseudoscalar glueball mass M with Wilson flow and fixed ending point, the
horizontal axis $r/a$ is the starting point of the preliminary fitting range. Left: $T=1.19T_c$,
right: $T=1.36T_c$¡£}
\end{figure*}

We find that when the starting point of the fitting range is fixed and the ending point is varied, once
the error bar of the TCDC at the ending point touches
the value zero, the fitting result is independent of the ending point. This phenomenon is also found
in Ref.~\cite{gbmass_fit1}. Therefore we fix the ending point that the error bar of the TCDC has touched
the value zero and vary the starting point
to extract preliminary pseudoscalar glueball mass $M$. Then we find the proper $\lambda$ and fitting window
to extract the final pseudoscalar glueball mass $M$. Results are showed in Fig.~\ref{fig:mass001}.

Both ensembles have the most stable plateau of the preliminary pseudoscalar glueball mass $M$ at $\lambda=0.16\textrm{fm}$.
Therefore we choose the data from $\lambda=0.16\textrm{fm}$ to extract $M$. The final fitting window is determined
by the range that the plateaus of the preliminary pseudoscalar glueball mass overlap with plateaus nearby.
In Fig.~\ref{fig:mass001}, red solid lines denote the final fitting results of the pseudoscalar glueball mass $M$, their
ranges represent the final fitting windows, pink dash lines represent the errors of the final pseudoscalar glueball mass $M$.
Numeric results are $T=1.19T_c,~M=1.915(98)\times 10^3\textrm{MeV}$ and $T=1.36T_c,~M=1.829(123)\times 10^3\textrm{MeV}$.
For comparing our results with those from Ref.~\cite{gbmass_fit2}, we had used same parameter $r_0 \approx 410\textrm{MeV}$
as Ref.~\cite{gbmass_fit2} does. The fitting results are consistent with those from Ref.~\cite{gbmass_fit2}. Noting that
the final fitting window in the left panel is shorter than that in the right panel. It should be owing to the coarser
lattice spacing $a$ of the ensemble in the left panel, same thing has also been found in Ref.~\cite{gbmass_fit1}. In fact,
we also apply the fitting procedure to ensembles at lower temperatures, which means ensembles with coarser lattice spacing $a$,
but fail to get proper final fitting windows to extract the final pseudoscalar glueball mass $M$. As for our
work, this method is available for extracting the pseudoscalar glueball mass at finite temperature with lattice spacing
$a < 0.08\textrm{fm}$.

\section{Summary}

In this paper we use Wilson flow to smear ensembles of quenched lattice QCD with lattice volume $32^3\times8$
at finite temperature. To study the topological structure of quenched QCD vacuum near $T_c$ corresponding to the critical inverse
coupling $\beta_c = 6.06173(49)$, we have used HS caloron-like topological lumps and IPR of
topological charge density. When the effective smearing radius $\lambda$ is large enough, we find
that the three quantities of HS caloron-like topological lumps are stable when $\beta \leq 6.050$. But 
these quantities change significantly when $\beta \geq 6.055$.
Similar behaviour is also found by using IPR to investigate the localization of topological charge density,
so the result is reliable. We extract the pseudoscalar glueball mass from TCDC at $T=1.19T_c,~
1.36T_c$, the results are consistent with those from conventional method. 
%Owing to the coarser lattice spacing $a$, this method is failed in the ensembles at lower temperatures .

\section*{Acknowledgments}

This work was mainly run on Tianhe-2 supercomputer at NSCC in Guangzhou. Supported in part by the
National Natural Science Foundation of China (NSFC) under the project No.11335001, No.11275169.
%\end{multicols}

\section{References}

\bibliographystyle{hunsrt}
\bibliography{ref}

\end{document}